\documentclass{emulateapj}
\usepackage{txfonts}
\usepackage{graphicx}
\usepackage{subfigure}
\usepackage{multirow}
\usepackage{url}

\shorttitle{Fast AO Image Selection in the Visible}
\shortauthors{N.M. Law et al.}

\begin{document}

\title{Getting Lucky with Adaptive Optics: \\Fast AO Image Selection in the Visible with a Large Telescope}

\author{N.M. Law\altaffilmark{1}, C.D. Mackay\altaffilmark{2}, R.G. Dekany\altaffilmark{1}, M. Ireland\altaffilmark{1,3}, J. P. Lloyd\altaffilmark{4}, A. M. Moore\altaffilmark{1}, J.G. Robertson\altaffilmark{3}, P. Tuthill\altaffilmark{3}, H. Woodruff\altaffilmark{3}}
\altaffiltext{1}{Department of Astronomy, Mail Code 105-24, California Institute of Technology, 1200 East California Blvd., Pasadena, CA 91125, USA; nlaw@astro.caltech.edu}
\altaffiltext{2}{Institute of Astronomy, University of Cambridge, Madingley Road, Cambridge, CB3 0HA, UK}
\altaffiltext{3}{School of Physics, University of Sydney, NSW 2006, Australia}
\altaffiltext{4}{Department of Astronomy, Cornell University, Ithaca, NY 14853}

\begin{abstract}

We describe the results from a new instrument which combines Lucky Imaging and Adaptive Optics to give the first routine direct diffraction-limited imaging in the visible on a 5m telescope. With fast image selection behind the Palomar AO system we obtained Strehl ratios of 5-20\% at 700 nm in a typical range of seeing conditions, with a median Strehl of approximately 12\% when 10\% of the input frames are selected. At wavelengths around 700 nm the system gave diffraction-limited 35 milliarcsecond FWHMs. At 950 nm the output Strehl ratio was as high as 36\% and at 500 nm the FWHM resolution was as small as 42 milliarcseconds, with a low Strehl ratio but resolution improved by factor of $\sim$20 compared to the prevailing seeing. To obtain wider fields we also used multiple Lucky-Imaging guide stars in a configuration similar to a ground layer adaptive optics system. With eight guide stars but very undersampled data we obtained 300 milliarcsecond resolution across a 30$\times$30 arcsec field of view in i' band.

\end{abstract}


\maketitle

\section{Introduction}

Adaptive optics has been used successfully for high-angular resolution imaging in the near-infrared on a large number of 5-10m class telescopes. However, it has not yet demonstrated routine diffraction-limited imaging in the visible on large telescopes. In this paper we describe the first system capable of producing diffraction-limited moderate-Strehl images in the visible on a 5m-class telescope. The system combines the Lucky Imaging \citep{Fried_1978, Baldwin_2001, Law_lucky_paper} frame selection concept and adaptive optics to produce much better performance than provided by either alone.

Direct Lucky Imaging has demonstrated routine diffraction-limited performance in I-band on 2.5m-class telescopes \citep{Mackay_2004_Lucky, Law_binaries_05, 2006MNRAS.368.1917L, Law_lucky_paper, Law_2008}. A simple, low-cost system has been developed by the Cambridge Lucky Imaging group and similar systems have now been deployed by other groups \citep[eg.][]{Hormuth_2007}. Lucky Imaging, in contrast to speckle imaging, provides direct images which can be used irrespective of the complexity of the science target.

\begin{figure*}
  \centering
  \resizebox{1.0\textwidth}{!}
   {
	\subfigure{\resizebox{0.9\columnwidth}{!}{\includegraphics[]{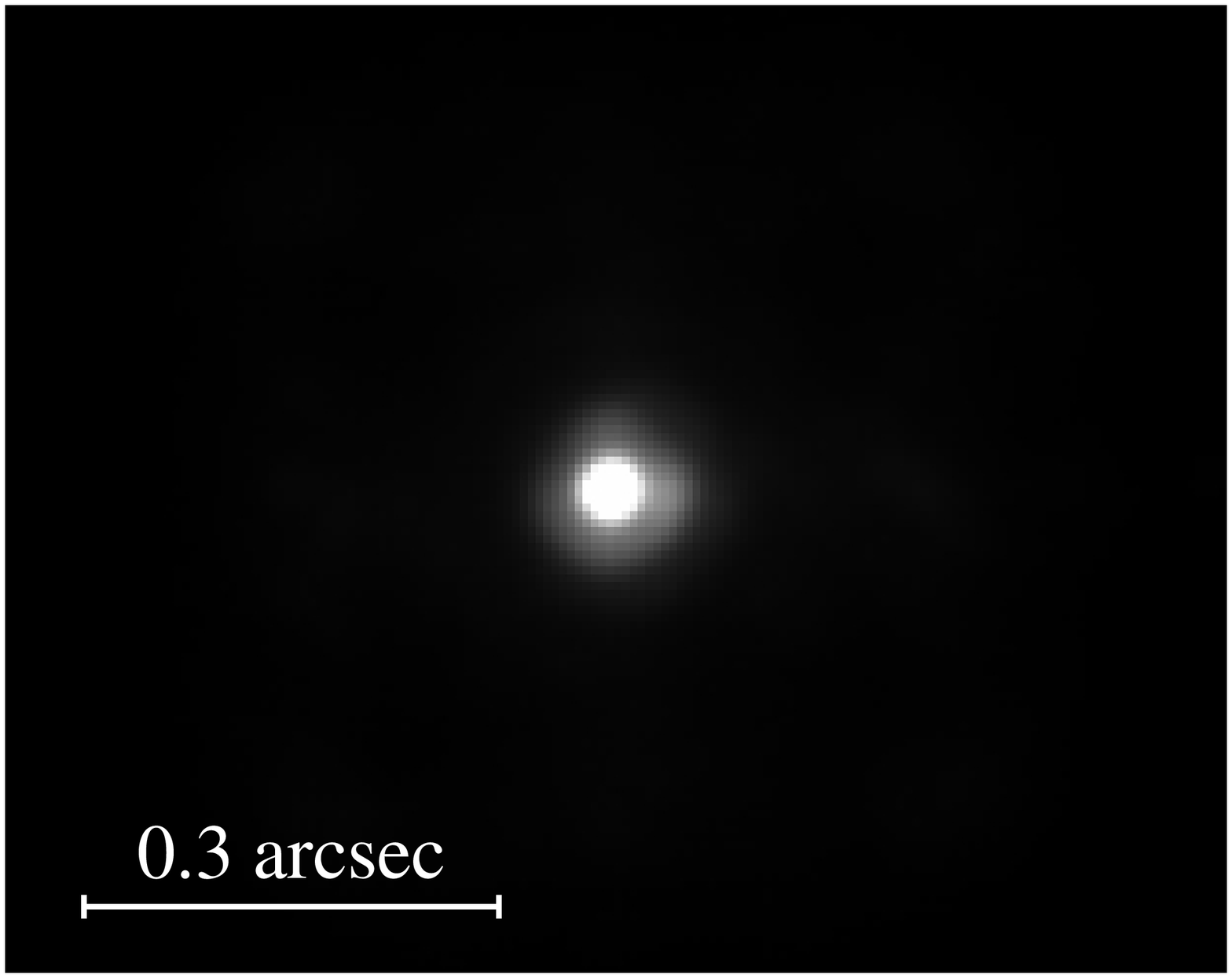}}}
	\hspace{0.3in}
	\subfigure{\resizebox{0.9\columnwidth}{!}{\includegraphics[]{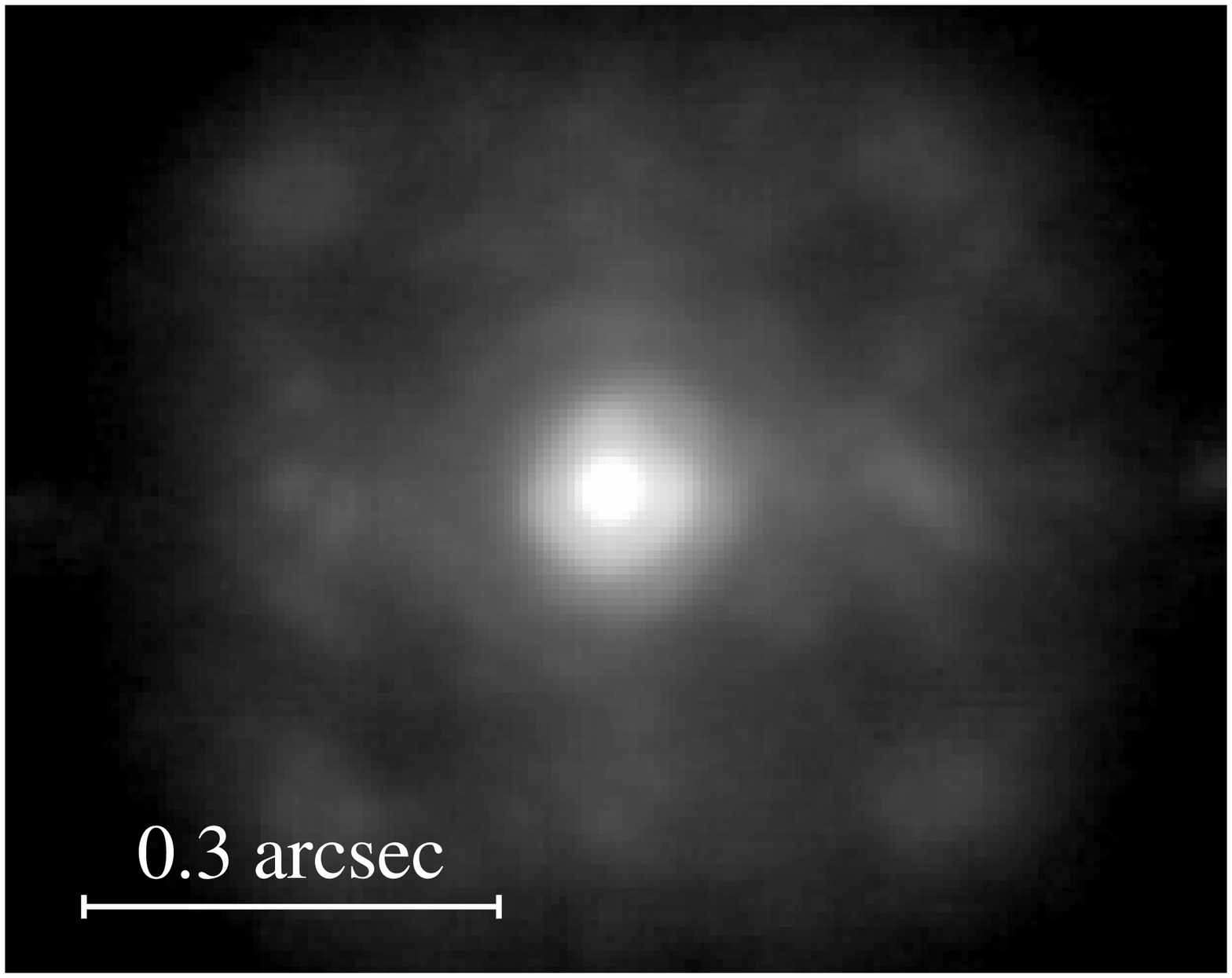}}}
	\hspace{0.3in}
	\subfigure{\resizebox{0.94\columnwidth}{!}{\includegraphics[]{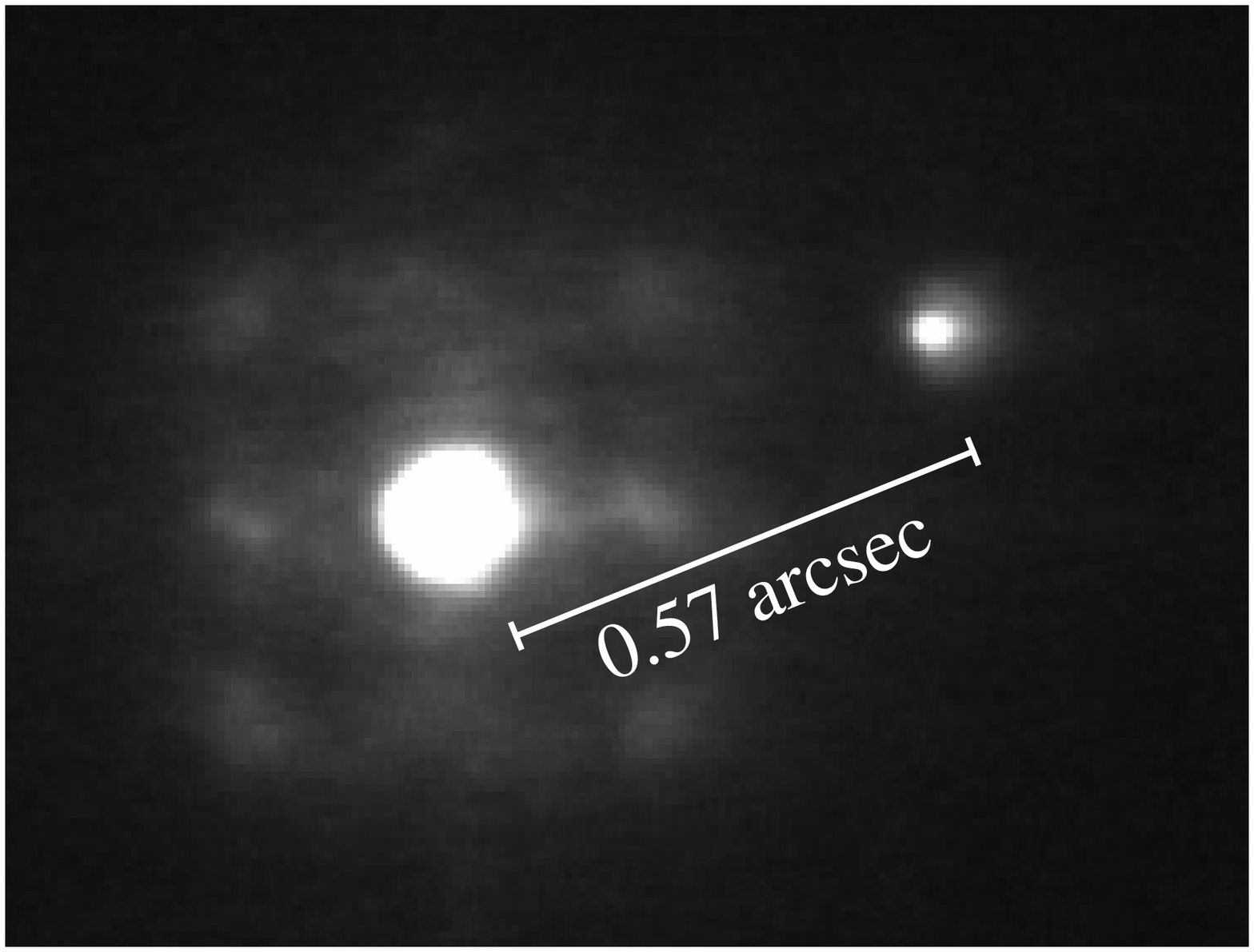}}}
   }
   \caption{Examples of Lucky+AO PSFs acquired with LAMP at 710 nm. \textit{Left:} The single star HD 160507; a 5\% selection from a 600 second run at 20 frames per second (linear grayscale). \textit{Middle:} the same image as the left panel but with a log scale. \textit{Right:} the binary HD 235089 ($\Delta m\sim4.5$, 0.57 arcsec separation) observed at 710nm (10\% selection from 320 seconds at 50 FPS).}
   \label{FIG:psf_examples}
\end{figure*}

We here describe the performance of visible-light Lucky Imaging behind an Adaptive Optics (AO) system on a large telescope (the experiment is hereafter referred to as Lucky+AO). We used the LAMP (LuckyCam, Adaptive optics, aperture Masking and polarimetry at Palomar) instrument, an electron multiplying (EM) CCD-based camera custom-built for this experiment to operate in conjunction with the Palomar Adaptive Optics system PALMAO. The instrument offered three observation modes: Lucky Imaging, adaptive-optics assisted aperture masking, and high-contrast imaging polarimetry. Results from the latter two modes will be described in separate publications.

During six nights on the Palomar 200" Hale telescope the system routinely produced diffraction-limited resolution images with 5-20\% Strehl at $\sim$700 nm (figure \ref{FIG:psf_examples}). In this paper we characterize the performance of the system under a variety of conditions and wavelengths with a view to guide the design of more permanent Lucky+AO instruments.

In section \ref{sec:obs} we describe the observations and instrument setup. In section \ref{sec:perf} we detail the performance of the system and the efficacy of the combination of Lucky Imaging and AO in the low-Strehl regime. We also describe our experiments in using multiple guide stars to obtain a larger useful field of view. We conclude in section \ref{sec:conc}.

\begin{figure*}
  \centering
  \resizebox{1.0\textwidth}{!}
   {
	\subfigure{\resizebox{0.9\columnwidth}{!}{\includegraphics[]{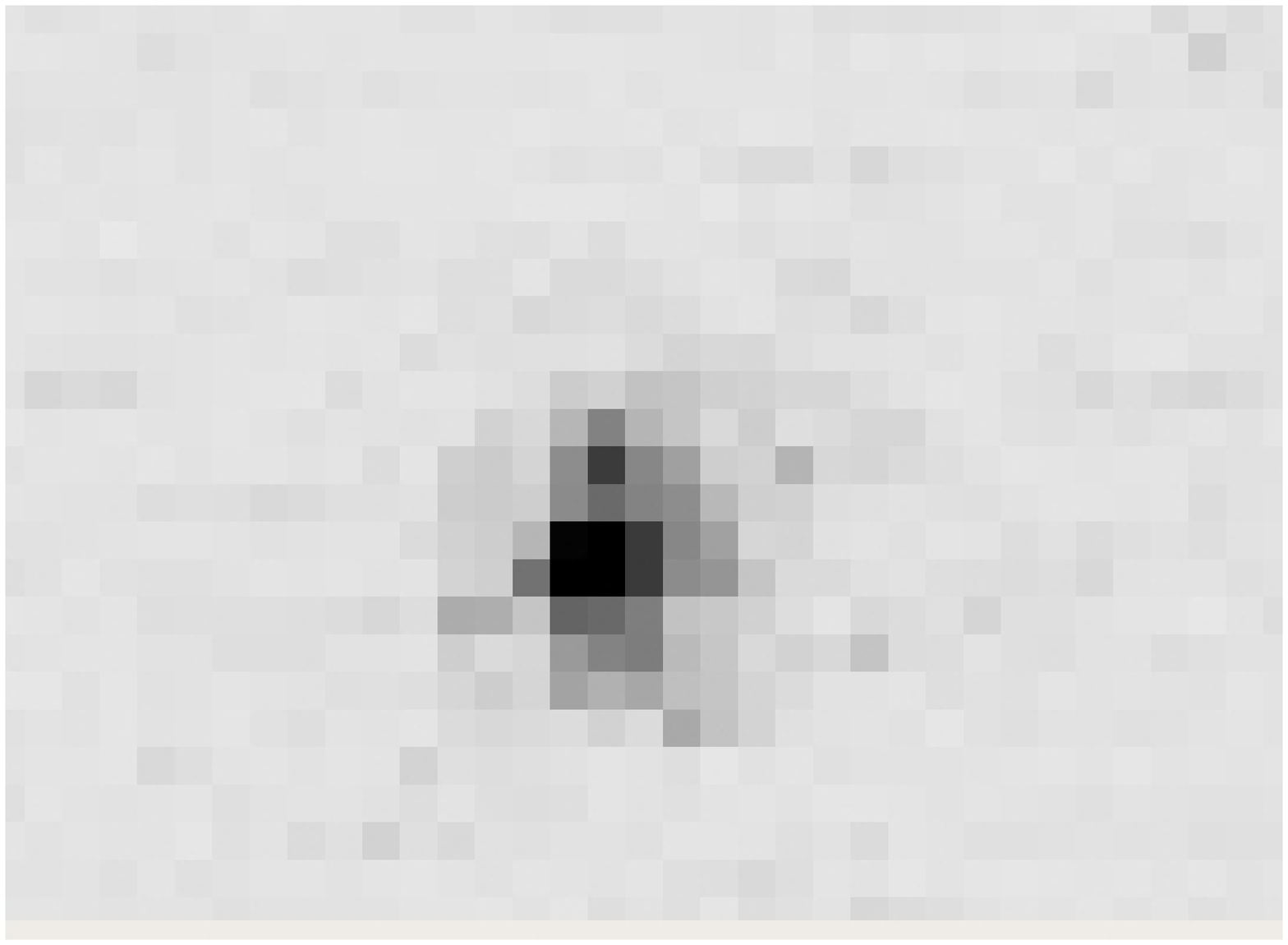}}}
	\hspace{0.3in}
	\subfigure{\resizebox{0.9\columnwidth}{!}{\includegraphics[]{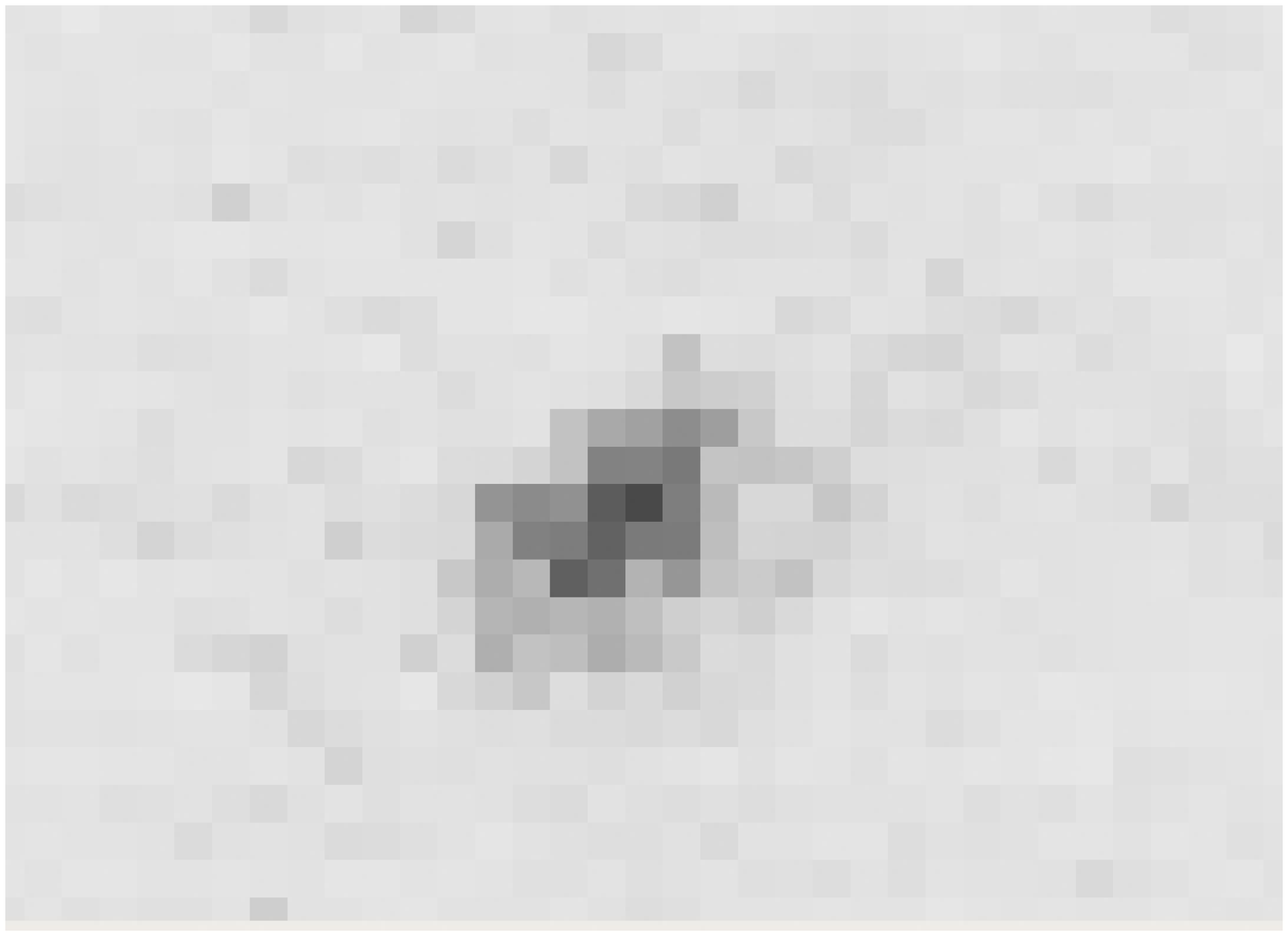}}}
	\hspace{0.3in}
	\subfigure{\resizebox{0.9\columnwidth}{!}{\includegraphics[]{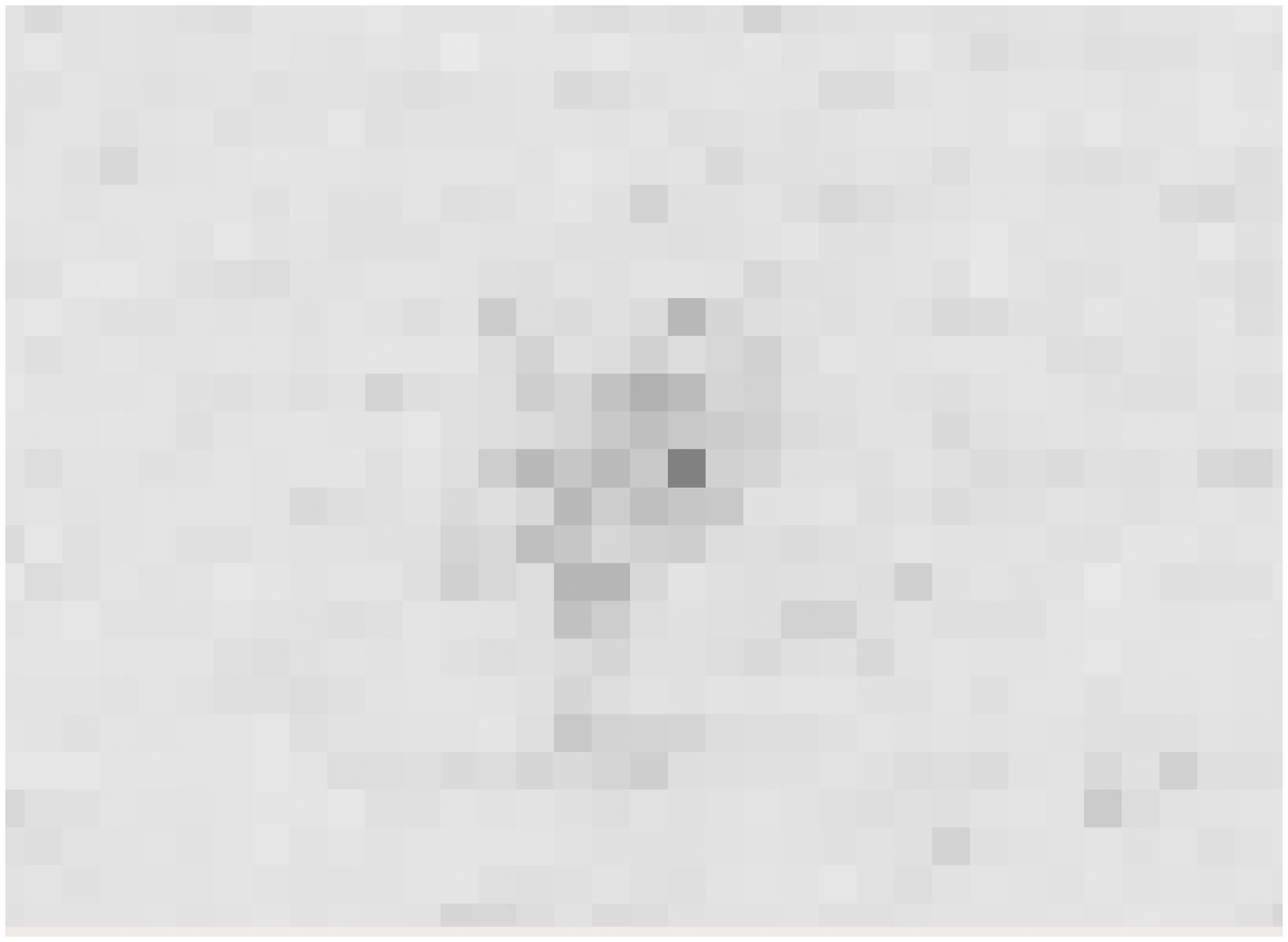}}}
   }
   \caption{Frames selected from 6000 frames in a typical Lucky Imaging + AO run taken at 50 FPS at 710nm with 14.9 milliarcsecond pixels, targeted at the single star HD 192849.  \textit{Left:} one of the high-Strehl frames in the run. \textit{Middle:} a frame with near-median Strehl. \textit{Right:} one of the lowest-Strehl frames.}
   \label{FIG:frame_comp}
\end{figure*}

\section{Observations and Instrument Setup}
\label{sec:obs}

LAMP was used in a six night run on the 5m Palomar Hale telescope from 2007 July 2 to 2007 July 8. Time equivalent to two nights was devoted to the Lucky Imaging mode described here, during which we observed a variety of both test and science targets.  The Palomar DIMM-MASS unit \citep{Thomsen_2007} was used to record seeing parameters during our observations.

LAMP was placed behind the Palomar Adaptive Optics system PALMAO \citep{Troy_2000, Dekany_1998}. The system has 241 active actuators with 5$\rm{\mu m}$ stroke, and a 16$\times$16 subaperture Shack-Hartmann wavefront sensor. The system operates at up to 2 kHz and typically updated at more than 200 Hz rates during our observations. In median Palomar conditions (V-band seeing = 1.1 arcsec) the typical bright-star wavefront error is 220nm RMS. We replaced the usual wavefront sensor visible/IR dichroic with a 50:50 beamsplitter to send visible light to the science focus.

LAMP was used in place of the AO system's usual near-infrared imaging camera. We built a simple reimaging camera to provide five field sizes ranging from 31 arcsec (61 milliarcsecond pixels, undersampled $\sim$4$\times$ at 700 nm) to 6.0 arcsec (Nyquist sampled at 500 nm). An atmospheric dispersion corrector was employed to allow broad-band observations. 

The camera detector was based on an electron-multiplying 528$\times$512 E2V CCD97. The CCD's electron-multiplication process allows detection of individual photons in each frame produced by the camera at the full quantum efficiency (up to 90\%) of the CCD \citep{Mackay_2004_L3CCD}. Our custom camera electronics are capable of running at up to 20 frames per second (FPS) in full 528x512 pixel frames; subarray readouts were used to increase the speed up to 50 FPS for some targets. The EMCCD gain was optimized for each target and ranged from 1 (no gain) to $\sim$10000. The camera produced 14-bit data at 7.5 megapixels per second. The data were recorded using custom software \citep{Law_thesis}; a lossless compression algorithm reduced the data transfer requirements by an average factor of 1.9, allowing direct recording onto external USB hard disks.

We used the standard Lucky Imaging data reduction pipeline without modification; complete descriptions of the reduction process can be found in \citet{Law_thesis} and \citet{Law_lucky_paper}. The data acquisition software was capable of displaying a realtime-preview of the Lucky Imaging output but the full data reduction was performed by a scripted process during daytime operations. Briefly, recorded frames were bias-corrected, flat-fielded and cosmic rays were removed. A bright star in the field was selected to serve as a Lucky Imaging guide star (in these observations this was typically the adaptive optics guide star). The frames were sorted in order of Strehl ratio and those that met a specific quality criterion were selected and aligned to produce a final high-angular-resolution image.

We estimate the Strehl ratio and optimal shift position for each image by cross-correlating the instantaneous guide star PSF with a diffraction-limited point spread function. The height of the peak of the resulting 2D array gives the degree of correlation ($\approx$ the Strehl ratio) and the position of the peak gives the optimal shift position to align the brightest speckles of each image. We perform the calculations on an image subsampled by 4$\times$4 to allow sub-pixel image alignment using the Drizzle \citep{Fruchter_2002} algorithm.

We targeted a variety of relatively bright ($\rm{m_V=6-10}$) stars for these observations to ensure that photon noise was a small contribution to any observed PSF variability (for our faintest observed stars we calculate that photon noise could contribute at most $\frac{1}{3}$ of the observed Strehl ratio variability).  Frame selection of images containing photon noise can bias FWHM measurements of the guide stars \citep{Law_thesis}; to avoid this we have measured all FWHMs and image profiles using the secondary component of close binaries.

\section{Performance}
\label{sec:perf}

\begin{figure}
  \centering
  \resizebox{1.0\columnwidth}{!}
   {
	\includegraphics{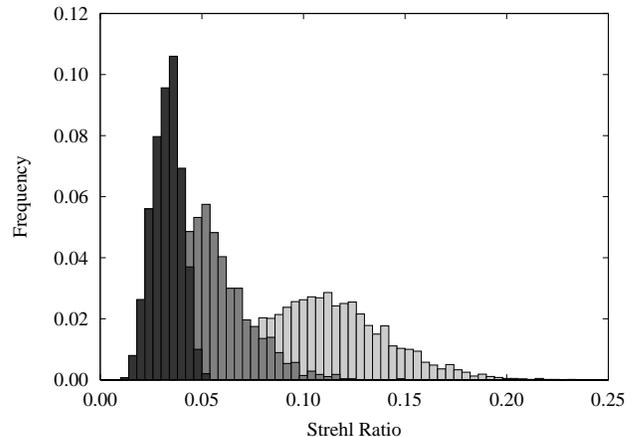}
   }
   \caption{Strehl ratio histograms for three 710nm wavelength observations from our dataset with seeings ranging from 0.75 to 1.1 arcsec. Note that for clarity the darker histograms obscure parts of the lighter ones.}
   \label{FIG:strehl_hists}
\end{figure}

Figure \ref{FIG:frame_comp} compares three representative frames recorded within a single 1-minute run. It is clear there are large image quality variations. Typical Strehl ratio histograms recorded by our instrument behind the AO system are shown in figure \ref{FIG:strehl_hists}. The histograms cover a representative range of seeings and frame rates for our dataset; it can be immediately seen that the Strehl ratio variations behind the AO system are sufficient that large Strehl ratio improvements can be realized by using only the frames with the greatest Strehl ratios.

Each of the Strehl ratio distributions is positively skewed, increasing the fraction of high-Strehl outliers compared with a Gaussian. The Strehl variability behavior in this low-to-moderate Strehl regime contrasts with the negatively-skewed distributions measured in the high-Strehl regime by \cite{Gladysz_2006, Gladysz_2008}. Our Strehl ratio distributions appear similar to the behavior observed for fast-frame-rate observations without an AO system \citep{Tubbs_2002, Law_thesis, Baldwin_2008}.  The Gladysz results were obtained in K-band on relatively small telescopes, with Strehl ratios up to 70\% and are thus in a very different image quality regime from our data. As noted in \cite{Gladysz_2006}, in the low Strehl regime covering the Lucky+AO experiment we would expect the observed positively-skewed Strehl ratio distributions.

\begin{figure}
  \centering
  \resizebox{1.0\columnwidth}{!}
   {
	\includegraphics{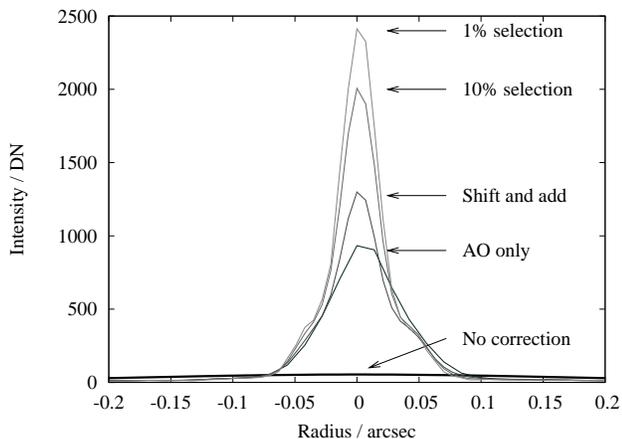}
   }
   \caption{Cuts through a typical Lucky Imaging + AO PSF. The companion to HD 235089 was observed in 0.86 arcsec seeing at 50 frames per second in a 10 nm bandpass centered at 710nm. The y-axis is in data numbers as recorded by the camera. The "no correction" (seeing-limited) profile is modeled here by a Gaussian with the width measured from the Palomar DIMM during the observations and total flux equal to that of our AO-corrected observations.}
   \label{FIG:psf_cuts}
\end{figure}

\subsection{Single Star PSFs}
\label{sec:psfs}

\begin{figure}
  \centering
  \resizebox{1.0\columnwidth}{!}
   {
	\includegraphics{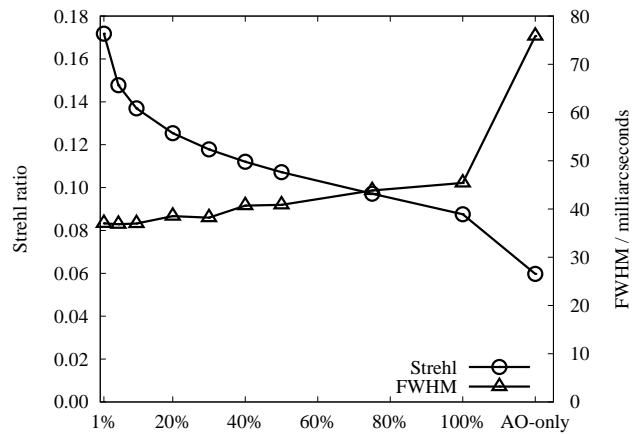}
   }
   \caption{Strehl ratio and full-width-at-half-maximum resolution as a function of frame selection percentage, for the same run as figure \ref{FIG:psf_cuts}. The run had a duration of 320 seconds during which the minute-timescale seeing was 0.86 arcsec with a peak-to-peak variability of 0.05 arcsec. AO-only is the result of a long-exposure (summing all the recorded frames) behind the AO system. 100\% "selection" is equivalent to shift-and-add using all the frames. Note that the data was slightly spatially undersampled, so the best resolution achievable for a diffraction-limited image is about 35 milliarcseconds.}
   \label{FIG:sel_effect}
\end{figure}

\begin{figure}
  \centering
  \resizebox{1.0\columnwidth}{!}
   {
	\includegraphics{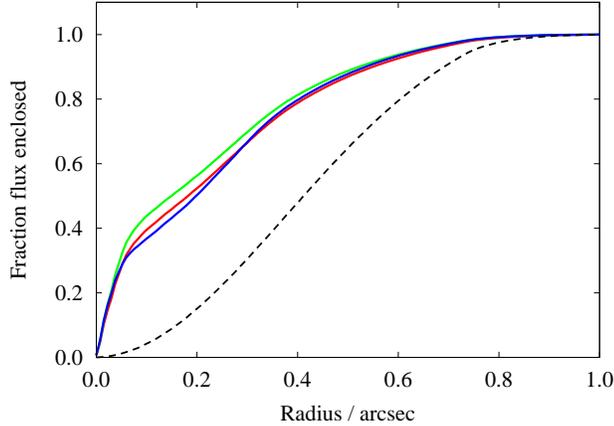}
   }
   \caption{The enclosed fraction of total starlight as a function of circular aperture radius. Three representative runs are shown (upper lines), all were 10\% Lucky+AO selections from runs on HD 192849 and HD 235089 taken at 50 FPS in a 10nm bandpass at 710nm; the seeing was 0.8-1.0 arcsec. The dashed line is a seeing-limited profile modeled by a Gaussian with 0.9 arcsec FWHM. Note that the structure of all the Lucky+AO curves is very similar, and the 50\% enclosed flux radius is less than 0.2 arcsec.}
   \label{FIG:enc_flux}
\end{figure}

Figure \ref{FIG:psf_cuts} shows cuts through the output images from a typical Lucky Imaging + AO run. The Airy ring only becomes visible in the PSF cuts after frame selection and alignment; frame selection leads to improved Strehl ratio and decreased FWHM. This is quantified for a typical run in figure \ref{FIG:sel_effect}. Using 20\% of the frames Lucky Imaging reached diffraction-limited resolution for the telescope and provided a $>$2$\times$ improvement in FWHM and Strehl ratio compared to using the AO system alone. Further improvements in Strehl ratio were achieved with more stringent frame selections while the FWHM remained diffraction-limited. Note that the more stringent selections concentrate light from the profile wings visible in figure \ref{FIG:psf_examples} and so appear above the unselected PSF at all radii shown in figure \ref{FIG:psf_cuts}. Figure \ref{FIG:enc_flux} details the enclosed flux as a function of circular aperture radius for three example Lucky+AO runs.

Under the conditions considered here the Strehl ratio increases rapidly as the quality criterion is made more stringent. The FWHM, however, is almost diffraction-limited even with only shift-and-add alignment and no frame selection (figure \ref{FIG:sel_effect}). Frame alignment reduces the FWHM in size by almost a factor of two, suggesting that a large component of the residual wavefront error produced by the AO system is in tip/tilt. However, we note that we guide on the brightest speckle rather than the centroid of the image, and so in the low-Strehl / multiple-speckle regime we obtain much higher quality PSFs than centroiding tip-tilt correction can provide \citep{Christou_1991}.

\begin{figure}
  \centering
  \resizebox{1.0\columnwidth}{!}
   {
	\includegraphics{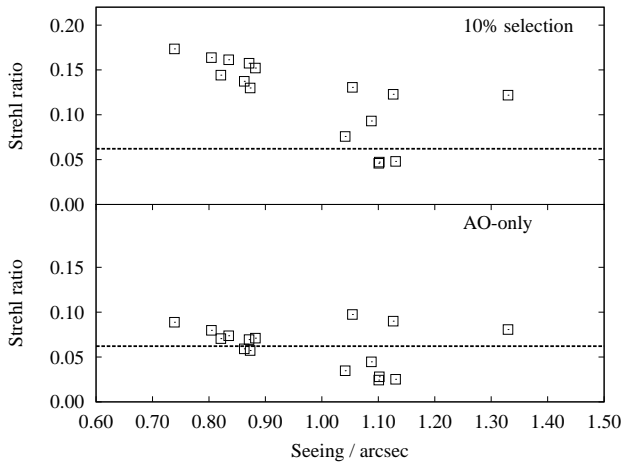}
   }
   \caption{Strehl ratio as a function of seeing. \textit{Bottom panel}: long exposure AO imaging (averaging the frames recorded by camera). \textit{Top panel}: The results of Lucky Imaging of the AO data, selecting 10\% of frames.  To guide the eye, the dashed line in each panel shows the average Strehl ratio achieved by the long-exposure AO imaging. To obtain a wide range of seeing the data presented here was taken over several hours on two separate nights. Each datapoint is a 1.0-4.0 minute dataset taken at frame rates between 20 and 50 FPS. Data was recorded in 10 nm passbands centered at 670 nm or 710nm; the images were sampled at 14.9 mas/pixel.}
   \label{FIG:seeing_strehl}
\end{figure}

Figure \ref{FIG:seeing_strehl} details the performance under a range of seeing conditions. In all observations Lucky+AO significantly improved on the performance of AO alone. Under median Palomar Observatory seeing (1.1 arcsec) and selecting 10\% of the frames Lucky+AO gave output Strehl ratios of 0.05 to 0.13 at 700 nm. In 0.7 arcsec seeing the performance increased to 17\% Strehl ratios. In all cases the system gave an improvement in Strehl of between 2$\times$ and 3$\times$ compared to the PALMAO system alone.

\vspace{1cm}
\subsection{Companion detection}

\begin{figure}
  \centering
  \resizebox{1.0\columnwidth}{!}
   {
	\includegraphics{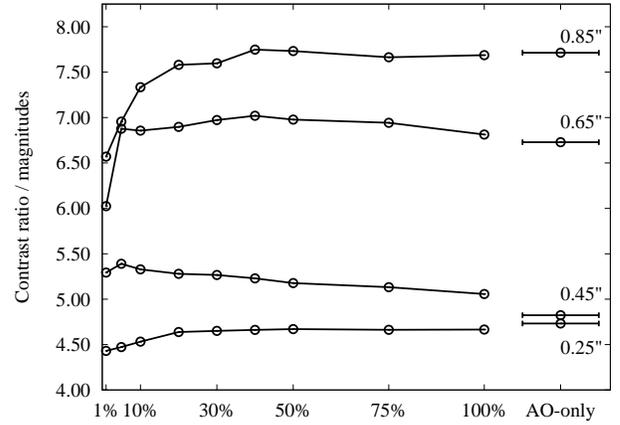}
   }
   \caption{Companion detection magnitudes. HD 235089, a 0.57" binary observed at 710nm, 50 FPS for 320 seconds is used to calculate the limiting contrast ratio at which a companion can be detected with a signal-to-noise ratio of 10. The specified binary separation is noted at the right of each line. The PSF of the real companion was used to determine the signal within a 6-pixel aperture. For each radius the RMS noise and its variance was measured at 8 apertures distributed around the primary. We conservatively set the effective background noise at that radius to be a 2$\sigma$ excursion above the average RMS noise. The bars at the right of the plot give the long-exposure performance behind the AO system, using the same data.}
   \label{FIG:conts}
\end{figure}

\begin{figure*}
  \centering
  \resizebox{1.0\textwidth}{!}
   {
	\subfigure{\resizebox{0.9\columnwidth}{!}{\includegraphics[]{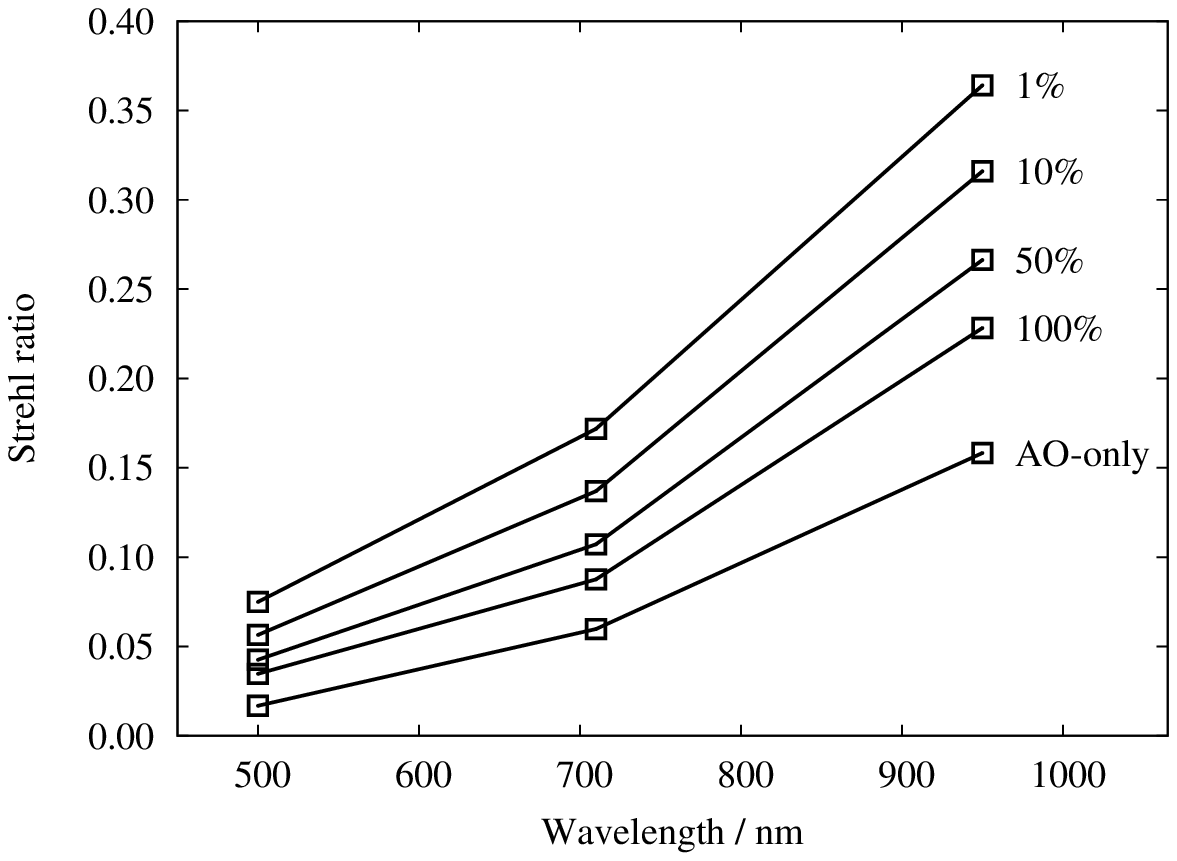}}}
	\hspace{0.3in}
	\subfigure{\resizebox{0.9\columnwidth}{!}{\includegraphics[]{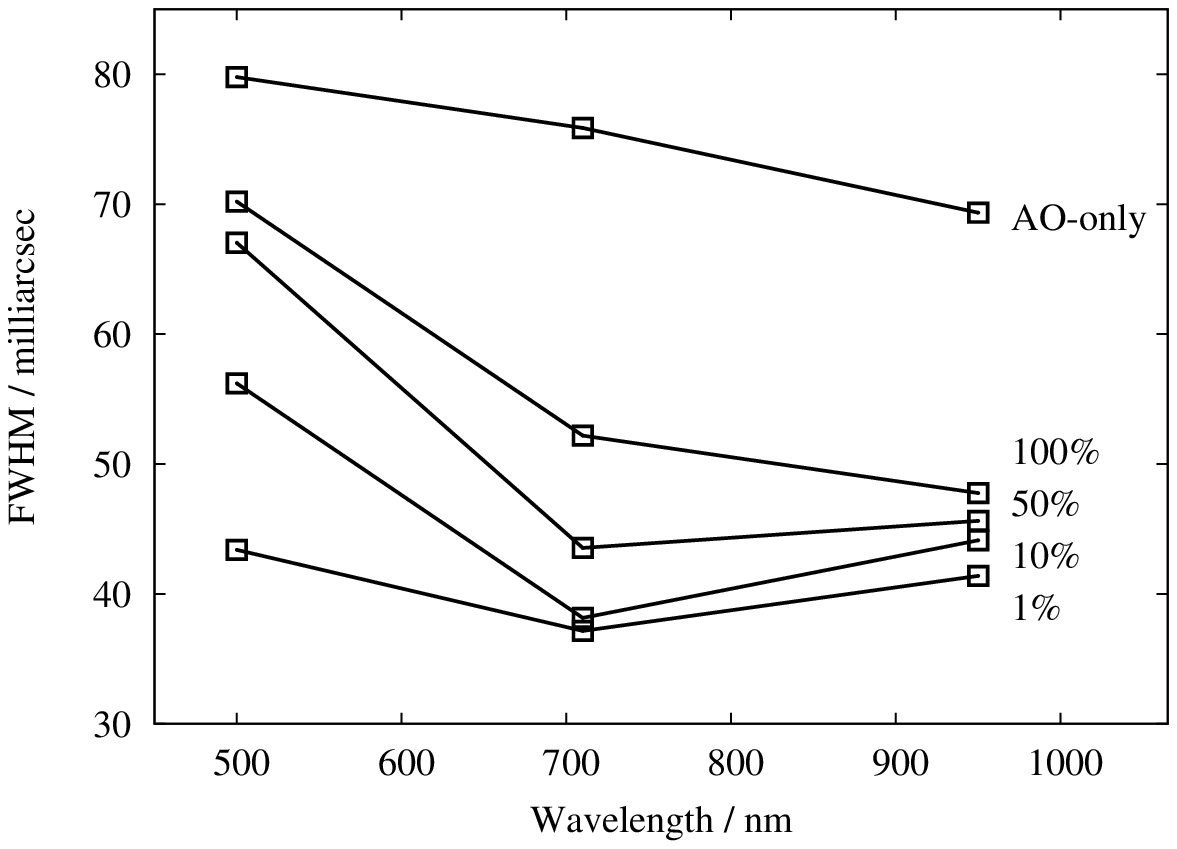}}}
   }
   \caption{Strehl ratio and FWHM as a function of wavelength and number of frames selected. Each observation was taken at 50 frames per second in a 10 nm bandpass filter. As in section \ref{sec:psfs}, to avoid biasing the FWHM measurement, the companion of HD 235089 was used as the performance measurement star while the AO and Lucky Imaging systems guided on the primary. Note that the increased FWHM of the Lucky-Imaging output PSFs at 950 nm compared to 700 nm is due to the increased diffraction-limited core size at those wavelengths. Slight undersampling of the original PSFs gives a slightly increased FWHM compared to that naively expected for a diffraction-limited PSF on a 5m telescope. 100\% selection is equivalent to shift-and-add brightest speckle alignment for all frames.}
   \label{FIG:wavelengths}
\end{figure*}

The benefits of the results described above are clear for crowded field data, where simply increasing the separation of stars through high angular resolution is most important. In this section we investigate the optimization of the Lucky Imaging process for faint, close stellar companion detection.

As the frame selection criterion is made more stringent the light in the PSF becomes more concentrated. This affects both the background against which a faint companion must be detected and the amplitude of its signal within a diffraction-limited core. However, the smaller effective integration time also increases the signal and background noise. The tradeoff between these effects is described in figure \ref{FIG:conts} where we show measured companion detection limits for a range of close companions vs. the frame selection level for a typical observation (based on the binary shown in the right panel of figure \ref{FIG:psf_examples}).

There is little or no drop in faint companion detection performance with frame selections as stringent as 10\%, at all tested radii. At the smallest separations ($\lesssim$0.1 arcsec, not shown), Lucky image selection is required to resolve the binaries. These results demonstrate that, compared to long-exposure AO imaging, Lucky+AO close companion detection can be performed with higher angular resolution and with minimal SNR performance loss. 

\subsection{Performance at other wavelengths}

To investigate the performance of Lucky+AO as a function of wavelength we observed HD 235089 in a set of bands between 500 and 950 nm (figure \ref{FIG:wavelengths}). The observations were taken during a 45-minute period during which the DIMM seeing was 0.80$\pm$0.15 arcsec. The wavelength range was limited in the blue by the lower throughput of the AO system optics, and in the red by our CCD detector's sensitivity. As above, we use the star's companion (0.6 arcsec separation) to avoid biasing the PSF shape by the influence of photon noise on frame selection.

In all cases the output FWHM resolution was improved by a factor of at least two compared to the capabilities of the AO system alone. At 500 nm the resolution achieved was as good as 43 milliarcseconds, not diffraction-limited but still a factor of $~$20 improvement over the natural seeing.

As would be expected, the system gives a decreased output Strehl ratio at shorter wavelengths. However, the \textit{fractional} increase in Strehl ratio is actually greater at 500 nm than at 710 nm, and the frame selection process gives the most obvious FWHM improvements at 500 nm. It seems that there is increased frame sharpness variance associated with the lower quality images at the shorter wavelengths. 

\subsection{Frame rates}

\begin{figure}
  \centering
  \resizebox{1.0\columnwidth}{!}
   {
	\includegraphics{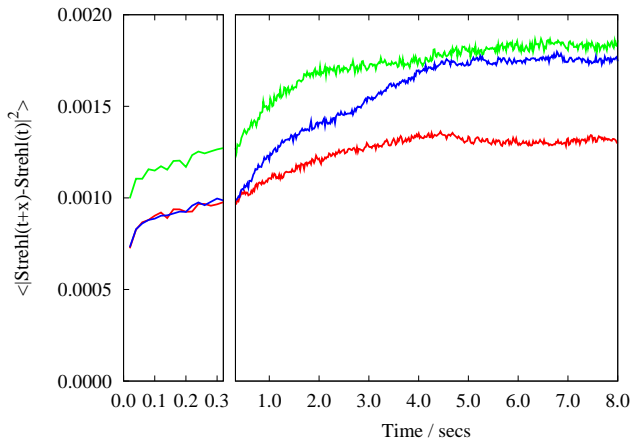}
   }
   \caption{Structure functions, the typical Strehl ratio variance between frames as a function of time, for three typical runs taken at 50 FPS at 710 nm in a 10 nm bandpass. Note the slight downturn at the minimum frame spacing of 0.02 seconds, suggesting that we are approaching but not reaching the coherence time.}
   \label{FIG:frame_rates}
\end{figure}

\begin{figure*}
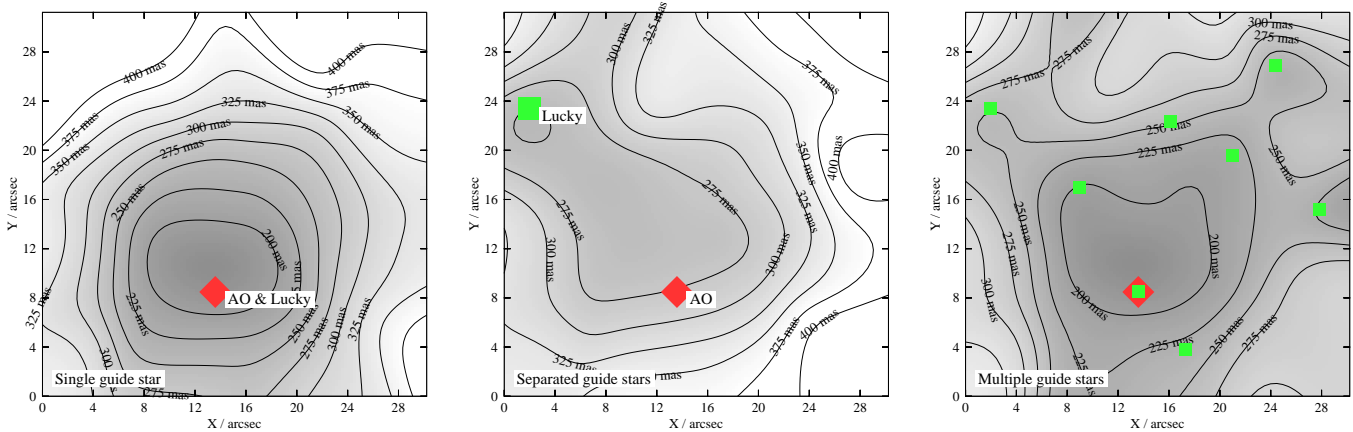

  \centering
  \resizebox{1.0\textwidth}{!}
   {
	\subfigure{\resizebox{0.32\textwidth}{!}{\includegraphics[]{plot_ao_lucky_same.ps}}}
	\hspace{0.15in}
	\subfigure{\resizebox{0.32\textwidth}{!}{\includegraphics[]{plot_top_left_star4.ps}}}
	\hspace{0.15in}
	\subfigure{\resizebox{0.32\textwidth}{!}{\includegraphics[]{plot_best_refs.ps}}}
   }
   \caption{Maps of the variation in FWHM over a 30-arcsecond field in the core of the M13 globular cluster, taken with severely undersampled pixels in the SDSS i' filter (770 nm central wavelength). The grayscale is set to the same levels in each image for comparison; the variation in FWHM is measured from the profiles of 74 stars distributed throughout the field. \textit{Left:} Lucky Imaging on the AO guide star. \textit{Middle:} Lucky Imaging on a different guide star. Note the change in the variation of the image resolution, becoming elongated between the Lucky Imaging and AO guide stars. \textit{Right:} an map constructed from the same data using 8 different Lucky Imaging guide stars simultaneously and selecting the optimal Lucky guide star for each point in the image.}
   \label{FIG:isoplan}
\end{figure*}

Figure \ref{FIG:frame_rates} gives Strehl ratio structure functions for three typical observations. It is clear from these measurements that at our highest speed of 50 FPS the instrument is not completely sampling the PSF changes behind the AO system, with most of the observed variation occurring on timescales on the level of single frames. For stars of the brightness used here photon and EMCCD multiplication noise accounts for at most $\frac{1}{3}$ of the observed single-frame-time variations in Strehl; it appears that we really are significantly undersampling the true behind-AO Strehl ratio variations. This might be expected given the hundreds of frames per second rates typically employed by the PALMAO system for turbulence correction. We note that future Lucky+AO systems may wish to provide 100+ FPS to achieve the highest possible Strehl ratios.

\subsection{Isoplanatic patch}

In the above sections we used the adaptive optics guide star as the Lucky Imaging guide star; the greater guide-star sensitivity of lucky imaging leads the AO guide star to always be bright enough for Lucky guiding. If there are multiple guide stars in the field, however, we can investigate more complex frame quality sensing strategies. In this experiment we use multiple guide stars in a 30-arcsecond field in the core of the M13 globular cluster.

To obtain a wide field of view we had to use very spatially and temporally undersampled data (60 milliarcsecond pixels, 20 FPS). The performance, while much better than the prevailing 0.60 arcsec seeing, is therefore greatly decreased compared to the results described earlier and what could be achieved with a faster, larger array camera system.

When operating the data reduction system in the usual mode of using the AO guide star as the Lucky Imaging reference the FWHM approximately doubles over a 15 arcsecond radius (figure \ref{FIG:isoplan}, left panel). A very different behavior is observed when a different star is used for Lucky Imaging guiding (figure \ref{FIG:isoplan}, middle panel). The performance around the Lucky imaging guide star is greatly improved, from 400 mas FWHM to 240 mas FWHM, while the performance around the AO guide star is reduced. Guiding on the new star caused different frames to be selected as well as different image shifts to be used; the system is not simply correcting tip/tilt anisoplanatism. It appears that the AO system occasionally "Luckily" produced corrections that improved the turbulence experienced in other parts of the image.

The possibility of correcting different areas of the field with different selections of the \textit{same} data suggests a method of improving the isoplanatic patch if multiple guide stars are available, in a setup similar to a multiple-guide-star ground layer adaptive optics (GLAO) system. Using eight lucky guide stars simultaneously, and selecting the optimal guide star to use for each point in the image, the entire 30$\times$30 arcsecond field can be processed to give FWHM resolutions better than 300 mas in I-band (figure \ref{FIG:isoplan} [right panel]).

\section{Discussion and Conclusions}
\label{sec:conc}

The combination of Lucky Imaging and Adaptive optics produced 35 milliarcsecond resolution at 700 nm on the Palomar 200" telescope, extending the useful wavelength range of the Palomar adaptive optics system into the visible. The performance detailed in the above sections is unlikely to be the best achievable with systems of this type. Our results suggest that such systems should run very rapidly, as even 50 FPS observations undersampled the Strehl ratio variations. In addition, some of our observations are of lower performance than would be expected for the prevailing seeing conditions (figure \ref{FIG:seeing_strehl}). This may be due to insufficiently precise non-common-path (NCP) error calibrations on some nights. Although the Palomar AO system has demonstrated NCP calibrations as precise as 34 nm wavefront error with the standard infrared camera, LAMP typically achieved a Strehl ratio of only $\sim$30\% on PALMAO's built-in turbulence-free white light source in the visible. Future Lucky+AO instruments will likely obtain higher performance as better NCP error calibration is achieved.

Although the observations described here were of relatively bright ($\rm{m_V = 6-10}$) stars, we note that the system is quite capable of guiding on much fainter stars. Lucky Imaging (without an AO system) can guide on stars as faint as $16^{\rm{th}}$ magnitude \citep{Law_lucky_paper} on 2.5m telescopes and the Lucky+AO technique is therefore capable of using any guide star that is bright enough for adaptive-optics operation (assuming broadband imaging and half the light going to the AO wavefront sensor). Laser guide star extensions to the Lucky+AO technique will require techniques to mitigate the extra background from the laser backscatter, but are possible in principle.

As the first system enabling diffraction-limited moderate-Strehl performance in the visible on 5m-class telescopes, the science applications are broad. In particular, crowded field visible photometry (for example in extragalactic resolved stellar population studies) and high-angular-resolution studies of nebular emission lines in the visible will benefit greatly from the increased resolution. 

The enlarged useful field of view possible with multiple guide stars may greatly improve the range and size of fields which can be covered. We recommend further investigation of the performance of this mode with higher frame rate cameras and improved PSF sampling.

The seeing during our experiments was typical for the Palomar Observatory site, mostly around one arcsecond. This offers the intriguing possibility that these resolutions could be improved using a larger telescope in better seeing conditions -- for example, the Keck telescope in 0.5 arcsec seeing may be capable of reaching its diffraction-limit in the visible using Lucky+AO techniques.

The combination of Lucky Imaging and AO offers a relatively simple, low-cost upgrade to current adaptive optics systems. As the first direct diffraction-limited imaging in the visible on large telescopes this technique pioneers the science which will be performed with future visible-AO systems.

\acknowledgments We particularly thank the PALMAO team, especially Jenny Roberts and Antonin Bouchez, for all their help during the design, setup and operation of the instrument. Thanks also go to the Palomar Observatory team for great assistance throughout our run. We thank Chris Koresko for providing the ADC design.

{\it Facilities:} \facility{Hale}

\bibliographystyle{apj}
\bibliography{lamp_perf}

\label{lastpage}

\end{document}